\documentclass[a4paper,11pt]{article}
\usepackage{graphicx,amssymb,bm,latexsym,color,epsf,inputenc}
\makeatletter
\@addtoreset{equation}{section}

\makeatother
\newcommand{\be}{\begin{equation}}
\newcommand{\ee}{\end{equation}}
\newcommand{\bea}{\begin{eqnarray}}
\newcommand{\eea}{\end{eqnarray}}
\newcommand{\ena}{\end{eqnarray}}
\newcommand{\vs}[1]{\vspace{#1 mm}}

\renewcommand{\b}{\beta}

\newcommand{\hk}{{\hat k}}

\newcommand{\nn}{\nonumber\\}

\newcommand{\cR}{{\cal R}}
\newcommand{\cD}{{\cal D}}
\newcommand{\cO}{{\cal O}}
\newcommand{\cW}{{\cal W}}

\newcommand{\bg}{\bar g}
\newcommand{\hg}{\hat{g}}

\newcommand{\bDelta}{\bar\Delta}
\newcommand{\bDeltaw}{\bar\Delta^W}

\newcommand{\hDelta}{\hat\Delta}

\newcommand{\bchi}{\bar\chi}
\newcommand{\bsigma}{{\bar\sigma}}
\newcommand{\bnabla}{\bar\nabla}

\newcommand{\hnabla}{\hat\nabla}
\newcommand{\tr}{{\rm tr}}
\newcommand{\Tr}{{\rm Tr}}
\newcommand{\Str}{{\rm STr}}

\newcommand{\Lie}{{\cal L}}
\newcommand{\deltaq}{\delta^{(Q)}}
\newcommand{\deltas}{\delta^{(S)}}
\newcommand{\deltab}{\delta^{(B)}}
\newcommand{\deltaf}{\delta^{(F)}}
\newcommand{\deltae}{\delta^{(E)}}
\def\mg{\mathbf{g}}
\def\mbg{\bar\mathbf{g}}
\def\mhg{\hat\mathbf{g}}

\def\mh{\mathbf{h}}
\def\mX{\mathbf{X}}

\begin{document}

\vs{10}
\begin{center}
{\Large\bf Split Weyl transformations in quantum gravity}
\vs{15}

{\large
Carlos M. Nieto\footnote{e-mail address: cnieto@sissa.it}$^{a,b,c}$
Roberto Percacci\footnote{e-mail address: percacci@sissa.it}$^{a,b}$
and Vedran Skrinjar\footnote{e-mail address: vskrin@sissa.it}$^{a,b}$
\vs{10}

$^a${\em International School for Advanced Studies, via Bonomea 265, I-34136 Trieste, Italy}

$^b${\em INFN, Sezione di Trieste, Italy}

$^c${\em Abdus Salam International Center for Theoretical Physics, 
Strada Costiera 11, 34151 Trieste, Italy}
}
%$^d${\em INFN, Sezione di Bologna, via Irnerio 46, I-40126 Bologna, Italy}

\vs{15}
%%%%%%%%%%%%%%%%%%%%%%%%%%%%%%%%
{\bf Abstract}
\end{center}
We discuss various realizations of the Weyl group
in the background field expansion of quantum gravity,
in the presence of a cutoff, as required in applications
of the functional renormalization group.
In order to study the background--dependence,
special attention is given to split gauge transformations,
which act on the background field and fluctuation
keeping the total metric unchanged. 
The results generalize previous works on global 
and local scale transformations.

%%%%%%%%%%%%%%%%%%%%%%%%%%%%%%%%
\section{Introduction}
%%%%%%%%%%%%%%%%%%%%%%%%%%%%%%%%

Almost all work on covariant quantum gravity is based on the background
field method. 
One begins by splitting the metric into background and quantum parts
\be
g_{\mu\nu}=\bg_{\mu\nu}+h_{\mu\nu}
\label{linpar}
\ee
and then performs a functional integral over $h_{\mu\nu}$.
In doing so one has to gauge-fix the invariance under diffeomorphisms.
It is very convenient to choose linear background gauges
of the form
\be
\label{gf1}
F_\mu=\bnabla_\rho h^\rho{}_\mu-\frac{\b+1}{d}\bnabla_\mu h^\rho{}_\rho\ .
%=\bnabla_\rho h^{T\rho}{}_\mu-\frac{\beta}{d}\bnabla_\mu h\ ,
\ee
The advantage of such gauges is that they 
break ``quantum'' diffeomorphisms
\be
\deltaq_v \bg_{\mu\nu}=\Lie_v g_{\mu\nu}\ ;\qquad
\deltaq_v h_{\mu\nu}=0
\ee
as required, while preserving background diffeomorphisms
\be
\deltab_v \bg_{\mu\nu}=\Lie_v \bg_{\mu\nu}\ ;\qquad
\deltab_v h_{\mu\nu}=\Lie_v h_{\mu\nu}\ .
\ee
The classical action, regarded as a functional of $\bg_{\mu\nu}$
and $h_{\mu\nu}$, is invariant under the ``split symmetry'' 
\be
\label{split}
\delta\bar g_{\mu\nu}=\epsilon_{\mu\nu}(x)\ ,\qquad
\delta h_{\mu\nu}=-\epsilon_{\mu\nu} (x)\ ,
\ee
simply because $g_{\mu\nu}$ is.
The gauge condition (\ref{gf1}) breaks this symmetry
and consequently the effective action is a functional of two
separate arguments $\Gamma(h;\bg)$.
This is not a very serious drawback, because one expects that 
the $n$-point functions of $h$ or $\bg$ lead to the same
physical results, once one goes on shell,
as is the case in YM theory \cite{Abbott:1983zw,Becchi:1999ir}.

The problem is more serious when one tries to calculate the
Effective Average Action (EAA) $\Gamma_k$, which is defined
by introducing in the functional integral a cutoff term
\be
\Delta S_k(h_{\mu\nu};\bg_{\mu\nu})=
\frac{1}{2}\int d^dx\sqrt{\bg}\,
h_{\mu\nu}\cR_k^{\mu\nu\rho\sigma}
h_{\rho\sigma}\ ,
\label{gencutoff}
\ee
where $\cR_k^{\mu\nu\rho\sigma}$ is constructed with the
background metric.
This introduces further breaking of the split symmetry,
which is not merely a gauge artifact
and spoils ``background-independence'',
the notion that physical results should not depend on
the choice of the background metric.

Is it possible to repackage the information contained
in the EAA into a functional of a single metric?
Pragmatically, much work on the renormalization group (RG) 
for gravity has concentrated on the functional 
$\bar\Gamma_k(\bg)=\Gamma_k(0;\bg)$
where one simply sets the classical fluctuation field to zero \cite{reuter1}.
By using the covariant Schwinger-DeWitt formalism,
one can sometimes compute beta functions without
specifying the background \cite{cpr2}, so that the result can be said
to be background-independent.

More recently there have been several calculations
of beta functions based either on ``bi-metric'' truncations
\cite{Manrique:2009uh}
or truncation depending on a flat background
and up to four powers of $h_{\mu\nu}$
\cite{donkin,cdp,dep,pawlowski}.
These calculations highlight the problem of the 
split symmetry breaking and raise the question
of how to physically interpret the results.
In the bi-metric case split-invariance has been imposed 
in the IR limit \cite{Becker:2014qya}.
Alternatively, one can try to solve simultaneously the modified
split symmetry Ward identity and the flow equation.
This was achieved in the conformally reduced case
\cite{dm3,Labus:2016lkh,Dietz:2016gzg}.
Other related ideas have been discussed in \cite{Safari:2016gtj,Wetterich:2016ewc}.

More progress has been made recently for the special case when
\be
\epsilon_{\mu\nu}=2\epsilon \bg_{\mu\nu}
\label{sst}
\ee
{\it i.e.} when the background is simply rescaled by a constant factor
\cite{Morris:2016spn,Percacci:2016arh,Ohta:2017dsq}.
In this case it was possible to write the anomalous Ward
identity explicitly. By making judicious choices
in the gauge-fixing and cutoff terms,
it has been reduced to the simple form
\be
\label{ward}
\delta_\epsilon\Gamma_k=\epsilon\partial_t\Gamma_k\ ,
\ee
where $t=\log k$ and the r.h.s. is just the
``beta functional'' of the theory.
The definition of the EAA contains a large
degree of arbitrariness, and in order to arrive at 
(\ref{ward}) one has to make several specific choices.
First and foremost, the splitting between the background and the
quantum field will not be of the standard linear form 
(\ref{linpar}) but rather of the exponential form,
see (\ref{exppar}) below.
Further specific choices have to be made in the gauge-fixing
and in the cutoff term.
In particular one has to use a ``pure'' cutoff, namely one 
that does not contain any running parameters 
\cite{narain2,Labus:2015ska}.
From the point of view of reducing the number of variables that the EAA depends on,
this relation can be used to eliminate
only the dependence on a single real degree of freedom,
namely the total volume of the background.

The main purpose of this paper is to study the generalization of
this result to the
case when the infinitesimal transformation parameter
$\epsilon$ in (\ref{sst}) is not a constant,
in other words when the background is subjected to a Weyl
transformation.
Because of the many different ways in which the Weyl group
can be realized in a physical theory,
it will be useful to pause to clarify the meaning of this statement.
The abstract Weyl group $\cW$ is just the multiplicative group
of positive real functions on a manifold.
It can be realized on fields in several ways
and to avoid the danger of misunderstandings
one should specify what realization one is talking about.
If the bare (classical) action is Weyl invariant,
then in addition to fixing the gauge for diffeomorphisms
one should also fix the Weyl gauge.
In this context, as with diffeomorphisms,
one will have to distinguish between
``quantum'' Weyl transformations
\be
\deltaq_\epsilon g_{\mu\nu}=2\epsilon g_{\mu\nu}\ ;\qquad
\deltaq_\epsilon \bg_{\mu\nu}=0\ ;\qquad
\deltaq_\epsilon h_{\mu\nu}=2\epsilon g_{\mu\nu}
\ee
and ``background'' Weyl (BW) transformations
\be
\deltab_\epsilon g_{\mu\nu}=2\epsilon g_{\mu\nu}\ ;\qquad
\deltab_\epsilon \bg_{\mu\nu}=2\epsilon \bg_{\mu\nu}\ ;\qquad
\deltab_\epsilon h_{\mu\nu}=2\epsilon h_{\mu\nu}\ .
\ee
What we will mostly be interested in here is a different realization,
which we shall call ``split Weyl transformations'' (SW)
\be
\label{bsc}
\deltas_\epsilon g_{\mu\nu}=0\ ;\qquad
\deltas_\epsilon \bg_{\mu\nu}=2\epsilon \bg_{\mu\nu}\ ;\qquad
\deltas_\epsilon h_{\mu\nu}=-2\epsilon \bg_{\mu\nu}\ .
\ee
Note that any bare action is invariant under (\ref{bsc})
simply because $g_{\mu\nu}$ is invariant under those transformations.
In fact, SW transformations are a subgroup of the split
transformation (\ref{split}).

A secondary aim of this paper is to highlight the relation between
certain results concerning the
fate of global and local scale transformations in quantum gravity.
Several results can be more easily discussed 
in the context of Conformally Reduced (CORE) gravity,
where only the spin-zero, conformal degree of freedom of the metric is dynamical.
In \cite{machado} the Functional Renormalization Group Equation (FRGE) was applied to CORE gravity
and it was noted that (a certain realization of) 
Weyl transformations could be either preserved or not,
depending on the choice of cutoff.
Subsequently, several studies have focused on SW
transformations in CORE gravity \cite{dm3}.
Our treatment will follow closely \cite{Percacci:2011uf}, 
where it was shown (albeit in a single-metric context)
how to maintain Weyl invariance in the functional RG.
With some changes, the results of that paper can be adapted to the present case.

The paper is organized as follows.
In Section 2 we define a cutoff for CORE gravity 
that leads to the modified SW Ward Identity (mSWWI) (\ref{ward}).
In Section 3 we discuss the choices (in particular in the gauge-fixing) that are needed to extend this result to full gravity.
In Section 4 contains a discussion of the results and their psosible
extensions.

%%%%%%%%%%%%%%%%%%%%%%%%
\section{CORE gravity}
%%%%%%%%%%%%%%%%%%%%%%%%

Several authors have considered CORE gravity as
an interesting theoretical toy model in which to test various ideas
related to the use of the ERGE.
This has been done both in the ``single field'' 
\cite{creh,machado}
and in the bi-field approximation 
\cite{Manrique:2009uh,dm3,Labus:2016lkh,Dietz:2016gzg}.

%%%%%%%%%%%%%%%%%%%%%%%%
\subsection{Definitions}

In CORE gravity one considers only metrics belonging to a
single conformal class.
Fixing a ``fiducial'', or reference metric $\hg_{\mu\nu}$ in this class,
every other metric can be obtained by a Weyl transformation
\be
g_{\mu\nu}=e^{2\sigma}\hg_{\mu\nu}.
\ee
Given any action $S(g)$, one obtains an action 
$S'(\sigma;\hg)=S(g(\sigma,\hg))$. 
Insofar as $\hg$ is kept fixed, the dependence on it is 
often not indicated.
In this way gravity is reduced to a scalar field theory.
For the field $\sigma$ one has an additive background-quantum split
\be
\label{confpar}
\sigma=\bsigma+\omega\ .
\ee
Thus, we can define a background metric 
\be
\bg_{\mu\nu}=e^{2\bsigma}\hg_{\mu\nu}, 
\ee
and the full metric is obtained from the background metric
by means of the Weyl transformation
\be
\label{coreb}
g_{\mu\nu}=e^{2\omega}\bg_{\mu\nu}\ .
\ee

Since we have three different metrics in total, there are several Weyl transformations we can perform in this setting. 
If the classical action is Weyl-invariant to begin with,
its CORE reduction is constant and the CORE theory is topological.
This is a somewhat trivial case, but one could still
discuss the fate of the transformations in the quantum theory.
Background Weyl transformations are defined by
\be
\deltab g_{\mu\nu}=2\epsilon g_{\mu\nu}\ ;\quad
\deltab\bg_{\mu\nu}=2\epsilon\bg_{\mu\nu}\ ;\quad
\deltab\hg_{\mu\nu}=0
\ee
and therefore
\be
\deltab \sigma=\epsilon\ ;\quad
\deltab\omega=0\ ;\quad
\deltab \bsigma=\epsilon\ .
\ee
For a generic gravitational action, its CORE reduction is
not constant, but is by construction invariant under the
SW transformations, which are defined by
\be
\deltas g_{\mu\nu}=0\ ;\quad
\deltas\bg_{\mu\nu}=2\epsilon\bg_{\mu\nu}\ ;\quad
\deltas\hg_{\mu\nu}=0
\ee
and therefore
\be
\label{deltas}
\deltas \sigma=0\ ;\quad
\deltas\omega=-\epsilon\ ;\quad
\deltas \bsigma=\epsilon\ .
\ee
One can define a third realization of the Weyl group,
acting on the fiducial metric in such a way as to maintain
the background (as well as the full) metric invariant.
For want of a better name, these transformations shall be called ``fiducial Weyl (FW) transformations'':
\be
\label{deltaf}
\deltaf g_{\mu\nu}=0\ ;\quad
\deltaf\bg_{\mu\nu}=0\ ;\quad
\deltaf\hg_{\mu\nu}=2\epsilon\hg_{\mu\nu}
\ee
and
\be
\label{deltaff}
\deltaf \sigma=-\epsilon\ ;\quad
\deltaf\omega=0\ ;\quad
\deltaf \bsigma=-\epsilon\ .
\ee
In CORE gravity one does not generally consider such transformations,
because the fiducial metric is kept fixed,
but we mention them here for later reference.

%%%%%%%%%%%%%%%%%%%%
\subsection{Cutoffs}

In CORE gravity, we introduce the cutoff in the functional 
integration over $\omega$.
It has the general form
\be
\label{cutoff}
\Delta S_k=\frac{1}{2}\int d^dx\sqrt{\bg^{\phantom a} }\,
\omega \cR_k(\bsigma,\hg) \omega.
\ee
The cutoff kernel $\cR_k$ is a function of a Laplace-type operator $\cO$
constructed with the fiducial metric
and the background conformal factor.
There are several choices for this operator,
including the use of Weyl-covariant derivatives. 
In total, we consider five types of cutoffs.
We will discuss two of them in this section, the remaining three
in Appendix \ref{Cutoffs}.

We start with the cutoff defined by using 
$\cO=\bDelta$ in (\ref{cutoff}), where $\bDelta=-\bg^{\mu\nu}\bar\nabla_{\mu}\bar\nabla_{\nu}$
is the Laplacian of the background metric.
For dimensional reasons, it can be written as
\be
\cR_{k}(\bDelta)=k^{d}r(y),
\label{firstcut}
\ee
where $r$ is a dimensionless function of the dimensionless variable $y=\bDelta/k^{2}$, that goes rapidly to zero for $y>1$ and tends to 1 for $y\to 0$.
The result of the transformations discussed in the previous section is    
\bea
    \delta^{(S)}\Delta S_k&=&-\frac{1}{2}\int d^{d}x\sqrt{\bg}(\epsilon \cR_{k}\omega+\omega \cR_{k}\epsilon)  \nn
    &&+\frac{1}{2}\int d^{d}x\sqrt{\bg}\omega\left[\epsilon d \cR_{k}+\epsilon\frac{\partial \cR_{k}}{\partial \bsigma}+\partial_{\mu}\epsilon\frac{\partial \cR_{k}}{\partial(\partial_{\mu}\bsigma)}+\mathellipsis\right]\omega\ ,\\
\delta^{(F)}\Delta S_k&=&0\ ,\\
\delta^{(B)}\Delta S_k&=&\frac{1}{2}\int d^{d}x\sqrt{\bg}\omega\left[\epsilon d \cR_{k}+\epsilon\frac{\partial \cR_{k}}{\partial \bsigma}+\partial_{\mu}\epsilon\frac{\partial \cR_{k}}{\partial(\partial_{\mu}\bsigma)}+\mathellipsis\right]\omega.
\eea
We clearly see that only $\delta^{(F)}$ provides a simple transformation rule for the cutoff, in fact, it is trivial. 
\footnote{For this reason, in \cite{machado},
where only transformations of the type $\deltaf$ were considered,
this was called a ``Weyl-preserving'' cutoff.}
This type of cutoff has also been used in \cite{dm3}
where the split Ward identity has been studied.
The infinite series of terms appearing in the other two expressions,
however, precludes deriving a useful expression for the mSWWI.
In \cite{dm3} this problem was circumvented by considering
only constant $\bar\sigma$,
in which case the terms involving $\partial\epsilon$
drop out and a manageable expression was obtained.

Here we shall try to avoid such restrictions on the fields
by introducing a SW-covariant derivative.
The general definition of Weyl-covariant derivatives
is given in Appendix \ref{weylcal}.
For the CORE case it is sufficient to note that 
$\partial_{\mu}\bsigma$ transforms
as a gauge field under $\deltas$.
Since the field $\omega$ transforms by a shift,
one can define its covariant derivative
$\cD_{\mu}\omega=\partial_{\mu}\omega+\partial_{\mu}\bsigma$.
It is {\it invariant} under $\deltas$,
so that the second covariant derivative
$\cD_\nu \cD_\mu\omega=\hat\nabla_\nu \cD_\mu\omega$
is also invariant and the Laplacian 
$\bDeltaw=-\bg^{\mu\nu}\cD_\mu \cD_\nu$
transforms simply as $\deltas\bDeltaw=-2\epsilon\bDeltaw$.

As we shall see in more detail below,
it will be useful to consider 
an ``extended'' transformation $\deltae$
which agrees with $\deltas$ on all fields but acts also
on the cutoff by
\be
\label{deltae}
\deltae k=-\epsilon k\ ,
\label{vark}
\ee
as dictated by dimensional analysis.
Thus, acting on any functional of the fields and $k$,
\be
\deltae=\deltas-\int d^dx\, \epsilon\,k\frac{\delta}{\delta k}\ .
\label{vare}
\ee
Note that since $\epsilon$ is generally not constant, we cannot assume
that $k$ is constant either. For the time being we take this just as a mathematical
fact and defer a discussion of its physical meaning until later.

The cutoff is now a function
\be
\cR_{k}(\bDeltaw)=k^{d}r(y), \quad\mathrm{with}\quad
y=\frac{1}{k^2}\bDeltaw\ .
\ee
The discussion of Appendix \ref{Exp}, where we consider the transformation of $r(y)\omega$ for $x$-dependent $k$, allows us to write the transformations $\deltaf$, $\deltab$ and $\deltas$ as follows
\bea
\label{Deltas}
\deltas\Delta S_k&=&\int d^dx\epsilon k\frac{\delta}{\delta k}\Delta S_k-\frac{1}{2}\int d^{d}x \sqrt{\bg}(\epsilon \cR_{k}\omega+\omega r_{0}\epsilon),\\
\deltaf\Delta S_k&=&-\frac{1}{2}\int d^{d}x\sqrt{\bg}\omega\left[\partial_{\mu}\epsilon\frac{\partial \cR_{k}\omega}{\partial(\partial_{\mu}\bsigma)}+\partial_\mu\partial_\nu\epsilon\frac{\partial \cR_{k}\omega}{\partial(\partial_\mu\partial_\nu\bsigma)}+\ldots\right],
\label{deltafw}
\\
\deltab\Delta S_k&=&\int d^dx\epsilon k\frac{\delta}{\delta k}\Delta S_k+\frac{1}{2}\int d^{d}x \sqrt{\bg}\omega k^{d}\sum_{n=1}^{\infty}r_{n}y^{n}\epsilon,
\eea

It will become clear later that transformations involving linear terms in $\omega$ do not contribute to the variation of the EAA, so they are harmless for the derivation of a Ward identity. On the other hand, the transformations involving the functional derivative with respect to $k$ lead to Ward identities with a known and compact form, as we will now show. 
We now derive the Ward identity associated to $\delta^{(S)}$ for this type of cutoff.

%%%%%%%%%%%%%%%%%%%%%%%%%%%%%%%%%%%%%%%%%%%%%%%%%%
\subsection{The modified Split-Weyl Ward Identity}

We start from the generating functional $W_k$, defined by
\be
e^{W_{k}(j;\bsigma,\hg)}=\int\mathcal{D}\omega\, 
e^{-S-\Delta S_k+\int j\omega}\ .
\ee
It is convenient to assume that $j$ is a scalar density,
to avoid the appearance of $\sqrt{\bg}$.
Taking into account that $S$ is invariant under $\deltas$,
the variation of $W_k$ is
\be
\deltas W_{k}(j;\hg,\bsigma)=
-\langle\deltas\Delta S_{k}\rangle
-\int d^dx j\epsilon\ .
\label{Deltaw}
\ee
From the definition of the EAA
\be
\Gamma_{k}(\langle\omega\rangle;\bsigma,\hg)=-W_{k}+\int d^{d}x j\langle\omega\rangle-\Delta S_k(\langle\omega\rangle)\ ,
\label{eaa}
 \ee
its transformation is
\be
\delta^{(S)}\Gamma_{k}=-\delta^{(S)}W_{k}
-\int d^{d}x j\epsilon
-\delta^{(S)}\Delta S_{k}(\langle\omega\rangle).
 \ee
The terms coming from the source cancel in the variation of $\Gamma_{k}$, and we end up just with 
\be
\deltas\Gamma_{k}=\langle\delta^{(S)}\Delta S_k\rangle-\delta^{(S)}\Delta S_k(\langle\omega\rangle).
\ee
Similarly, the linear terms in $\omega$ coming from $\delta^{(S)}\Delta S_k$ cancel out, and we find
\be
\delta^{(S)}\Gamma_{k}=
\frac{1}{2}\Tr\left(\frac{\delta^{2}\Gamma_{k}}{\delta\omega\delta\omega}+\cR_{k}\right)^{-1}\int d^dx\epsilon k\frac{\delta \cR_{k}}{\delta k}\ ,
\label{Trace}
\ee
where we have used the relation $\left(\frac{\delta^{2}\Gamma_{k}}{\delta\omega\delta\omega}+\cR_{k}\right)^{-1}=
\langle\omega(x)\omega(y)\rangle
-\langle\omega(x)\rangle\langle\omega(y)\rangle$. 
Eq. (\ref{Trace}) tells us that the split symmetry in $S$ is broken at the quantum level due to the introduction of the cutoff action. On the other hand, the result of Appendix \ref{localap} tells us that the effective action, for an $x$-dependent scale, satisfies the flow equation 
\be
\int d^dx\delta k\frac{\delta\Gamma_{k}}{\delta k}=
\frac{1}{2}\Tr\left(\frac{\delta^{2}\Gamma_{k}}{\delta\omega\delta\omega}+\cR_{k}\right)^{-1}\int d^dx\delta k\frac{\delta \cR_{k}}{\delta k}\ .
\ee
Therefore, the variation of the effective action with respect to the transformation $\delta^{(S)}$ is proportional to the functional derivative with respect to the scale $k$
 \be
     \delta^{(S)}\Gamma_{k}=\int d^dx\epsilon k\frac{\delta\Gamma_{k}}{\delta k}\ .
 \ee
This last expression states that $\Gamma_{k}$ is invariant under 
the extended transformation $\deltae$ defined in the previous section:
\be
\deltae\Gamma_{k}=0\ .
\ee
The EAA can thus be written in terms of the invariant quantities $\hat k=e^{\bsigma}k$ and $\sigma=\bsigma+\langle\omega\rangle$ as
\be
\Gamma_{k}(\omega;\bsigma,\hg)=\hat\Gamma_{\hat k}(\sigma;\hg).
\label{COREmSWWI}
\ee
In this way we have been able to reduce by one the number of functions that the EAA depends upon.  
We will now try to extend this result to the case of full gravity.

%%%%%%%%%%%%%%%%%%%%%%%%
\section{Full gravity}\label{Full}
%%%%%%%%%%%%%%%%%%%%%%%%

%%%%%%%%%%%%%%%%%%%%%%%%
\subsection{Variations}
%%%%%%%%%%%%%%%%%%%%%%%%

The discussion of CORE gravity makes it clear that the
exponential parametrization of the conformal factor
is the most natural one, because Weyl transformations then
act additively on the field.
This suggests that it may be convenient
to parametrize the full metric fluctuation exponentially
rather than additively as in (\ref{linpar})
\cite{Kawai:1989yh,Eichhorn:2013xr,pv1,nink,Codello:2014wfa,
Gies:2015tca,pereiraI,pereiraII,Eichhorn:2015bna,Labus:2015ska,
opv,Dona:2015tnf}.
Let us therefore write
\be
g_{\mu\nu}= \bg_{\mu\rho}(e^\mX)^\rho{}_\nu
\quad
\mathrm{where}
\quad
\label{exppar}
X^\rho{}_\nu=\bg^{\rho\sigma}h_{\sigma\nu}\ .
\ee

In dealing with exponentials, it is convenient to suppress indices and treat
two-index tensors as matrices, independent of the position of the indices.
Thus (\ref{exppar}) will be written
$\mg=\mbg e^\mX$ and $\mX=\mbg^{-1}\mh$.
We decompose the fluctuation field into its tracefree and trace parts:
\be
\mX=\mX^T+2\omega\mathbf{1}
\label{decomp}
\ee
where $\mX^T$ is traceless and (following \cite{pv1})
we have defined $\omega=h/2d$, with $h=\tr\mX=\bg^{\mu\nu} h_{\mu\nu}$.
Then we can write
\be
\mg= \mbg\,  e^{2\omega} e^{\mX^T}\ ,
\label{alice}
\ee
which is the obvious generalization of (\ref{coreb}).
If the background metric undergoes the finite transformation
$\mbg\to\mbg e^{2\epsilon}$, invariance of the full metric
can be maintained by the compensating transformation
$\omega\to\omega-\epsilon$, while $\mX^T$ is left invariant.
Then $\deltas\mh^T=\deltas(\mbg \mX^T)=2\epsilon\mbg X^T=2\epsilon\mh^T$,
which implies that 
\be
\deltas h^{T\mu}{}_\nu=0\ ,\qquad
\deltas h^T_{\mu\nu}=2\epsilon h^T_{\mu\nu}\ ,\qquad
\deltas \omega=-\epsilon\ .
\label{ennio}
\ee

In this paper we will consider background metrics belonging to a single
conformal equivalence class.
As in the CORE case, we choose a fiducial metric $\hg_{\mu\nu}$
in this class and parametrize all the others by their conformal factor
$e^\bsigma$:
\be
\mbg=\mhg e^{2\bsigma}\ .
\label{bgexp}
\ee
Note that we can then write
\footnote{A somewhat similar splitting in the gravitational path integral has been advocated in \cite{thooft}.}
\be
\mg= \mhg\,  e^{2\sigma} e^{\mX^T}=\mbg\,e^{2\omega}e^{\mX^T}\ ;\qquad
\mbg= \mhg\,  e^{2\bsigma} 
\label{bob}
\ee
where $e^{\sigma}$ is the conformal factor 
of the full metric, which can be split into a background part $e^{\bsigma}$ and a quantum part $e^{\omega}$,
related again as in (\ref{confpar}). 
The basic transformation rule $\deltas\mbg=2\epsilon\mbg$ implies 
\be
\deltas\bsigma=\epsilon
\label{varbsigma}
\ee 
and this together with (\ref{ennio})
implies $\deltas\sigma=0$. Thus with these definitions the invariance
of the full metric can be expressed again as a simple shift invariance,
albeit of the arguments of exponentials.

%%%%%%%%%%%%%%%%%%%%%
\subsection{Gauge fixing}
%%%%%%%%%%%%%%%%%%%%%

We can use the Weyl calculus explained in Appendix to write
SW-invariant functionals.
Let us consider a gauge fixing term
\be
\label{gfaction}
S_{GF}=\frac{1}{2\alpha}\int d^d x \sqrt{\bg}\,
F_\mu Y^{\mu\nu} F_\nu\ ,
\ee
where $F_\mu$ is of the form (\ref{gf1}).
As shown in \cite{Percacci:2016arh}, $F_\mu$ is invariant under
{\it global} SW transformations, {\it i.e.}
transformations (\ref{sst}) with constant $\epsilon$.
It is easy to extend this result to {\it local} transformations
simply replacing the derivative $\bnabla$ by the Weyl-covariant 
derivative $\cD$ defined in (\ref{wcovder}).
Thus the gauge condition is now
\be
\label{gf2}
F_\mu=\cD_\rho h^\rho{}_\mu-2(\b+1)\cD_\mu \omega\ .
\ee
(we recall that $\omega=h/2d$) and we have
\be
\deltas F_\mu=0\ .
\ee

In \cite{Percacci:2016arh} invariance of the gauge-fixing action was
obtained by choosing $Y^{\mu\nu}$ to contain a power of the background Laplacian.
Here we note that the covariant derivative $\cD$ has separate dependences
on $\bg_{\mu\nu}$ and $\bsigma$, {\it i.e.} it cannot be written
in terms of $\bg_{\mu\nu}$ alone.
Thus the gauge-fixing action is a functional 
$S_{GF}(h^{T\mu}{}_\nu,\omega;\bg_{\mu\nu},\bsigma)$.
Given that there is already a separate dependence on $\bsigma$,
aside from the dependence through $\bg_{\mu\nu}$,
we may as well use it to define
\be
Y^{\mu\nu}=e^{-(d-2)\bsigma}\bg^{\mu\nu}\ .
\ee
This makes the gauge-fixing action invariant,
without introducing additional derivatives.
In particular, there is no need to introduce an auxiliary 
Nielsen-Kallosh ghost.

In order to derive the Faddeev-Popov operator, we start from
the transformation of the full 
metric under an infinitesimal diffeomorphism $\eta$,
$\delta_\eta \mg=\Lie_\eta \mg$.
The ``quantum'' gauge transformation of the background $\mbg$
and fluctuation field $\mX$ satisfy
\be
\delta^{(Q)}_\eta \mbg=0\ ;
\qquad
e^{-\mX}\delta^{(Q)}_\eta e^\mX
=e^{-\mX}\mbg^{-1}\Lie_\eta \mg
=e^{-\mX}\mbg^{-1}\Lie_\eta\mbg e^\mX
+e^{-\mX}\Lie_\eta e^\mX
\ .
\ee
Under any variation $\delta$,
$e^{-\mX}\delta e^\mX =\frac{1-e^{-ad_\mX}}{ad_\mX}\delta \mX$,
so using this on both sides we obtain
\be
\delta^{(Q)}_\eta\mX=
\frac{ad_\mX}{e^{ad_\mX}-\mathbf{1}}
\mbg^{-1}\Lie_\eta\mbg
+\Lie_\eta \mX\ .
\ee
The Faddeev-Popov operator, acting on a ghost field $C^\mu$, 
is defined by
\be
\Delta_{FP\mu\nu} C^\nu=
\cD_\rho\left(
(\delta^{(Q)}_C \mX)^\rho{}_\mu
-\frac{1+\beta}{d}\delta^\rho{}_\mu
\tr(\delta^{(Q)}_C \mX)
\right)
\label{elena}
\ee
where the infinitesimal transformation parameter $\eta$
has been replaced by the ghost $C^\mu$.
The full ghost action then has the form \cite{pereiraII}
\be
S_{gh}(C^*_\mu,C_\mu;\bg_{\mu\nu},\bsigma)=
-\int d^dx\sqrt{\bg}\,
C^*_\mu Y^{\mu\nu}\Delta_{FP\nu\rho} C^\rho \ .
\ee
The infinitesimal diffeomorphism parameter $\eta^\mu$,
and hence the ghost field $C^\mu$, can be assumed to
be invariant under $\deltas$.
Then, a straightforward calculation shows that
$\delta^{(Q)}_C\mX$ is invariant.
Consequently, also $\Delta_{FP\mu}{}^\nu C_\nu$ is invariant.
Assuming that the antighost $C^*_\mu$ is also invariant,
the transformation of $Y^{\mu\nu}$ then exactly cancels the transformation
of the integration measure, and we conclude that $S_{gh}$ is
SW-invariant.
\footnote{These transformation of the ghost $C_\mu$ and antighost $C^*_\mu$ agree with those of \cite{Percacci:2016arh}
when $\epsilon$ is constant.}
Note that this statement refers to the full ghost action,
containing infinitely many interaction vertices that are bilinear
in the ghosts and contain arbitrary powers of $h_{\mu\nu}$.

%%%%%%%%%%%%%%%%%%%%%%%%
\subsection{Cutoff terms}
%%%%%%%%%%%%%%%%%%%%%%%%

We now have to generalize the cutoff choice discussed in \cite{Percacci:2016arh},
from constant to non-constant rescalings of the metric.
As before, it will be useful to consider also transformations
where the cutoff itself changes (see Eq. (\ref{vark})).
Our ultimate goal is to arrive at an EAA that is invariant under the 
extended transformations, and the way to achieve it is to concoct the
cutoff term in such a way that it is itself invariant.
This issue has been addressed previously in
a slightly different context in \cite{Percacci:2011uf,Codello:2012sn}.
Here we shall use the same techniques to write cutoff actions that are
invariant, except for a single term that has to do with the inhomogeneous transformation properties of $\omega$.
We shall see that this is not an obstacle for the construction
of an invariant EAA.

In order to construct diffeomorphism- and Weyl-invariant cutoffs
we use a Weyl-covariant second order differential operator. 
For definiteness we adopt a ``type I'' cutoff
(in the terminology of \cite{cpr2}) depending on the Laplacian
\be
\bDeltaw=-\bg^{\mu\nu}\cD_\mu\cD_\nu\ .
\label{blap}
\ee
The cutoff terms for all the fields have the structure
\bea
\Delta S_k^T(h^T;\bg,\bsigma)&=&
\frac{1}{2}\int d^dx\sqrt{\bg}\,
h^{T\mu}{}_\nu\cR_k(\bDeltaw) h^{T\nu}{}_\mu\ ,
\nonumber\\
\Delta S_k^\omega(\omega;\bg,\bsigma)&=&
\frac{1}{2}\int d^dx\sqrt{\bg}\,\omega\,\cR_k(\bDeltaw)\omega\ ,
\nonumber\\
\Delta S^{gh}_k(C^*,C;\bg,\bsigma)&=&
\int d^dx\sqrt{\bg}\,
C_\mu^* \cR_k(\bDeltaw) C^\mu\ ,
\label{allgencutoff}
\eea
where 
\be
\cR_k(\bDeltaw)=\,k^d r(y)
\ , 
\qquad y=\frac{1}{k^2}\bDeltaw.
\label{gencutoff2}
\ee
We have chosen the cutoff terms to be diagonal in field space,
without loss of generality.
Except for the introduction of the
Weyl-covariant derivatives, these cutoffs
are the same as in \cite{Percacci:2016arh}.

Note that we write the cutoff in terms of the mixed fluctuation so that all the fields have weight zero, i.e., they are invariant, except
for $\omega$ that transforms by a shift.
For a general tensor of weight $\alpha$, the operator $\bDelta^W$ generates a tensor of weight $\alpha-2$.
Thus we can write
\be
\deltae\bDelta^W=-2\epsilon\bDelta^W+\alpha[\epsilon,\bDelta^W]\ .
\ee
This implies that $r(y)$ maps a tensor of weight $\alpha$ 
to another tensor of weight $\alpha$ under $\deltae$.
Therefore, by simple counting, the cutoff terms for $h^{T}$ and $C$ are 
invariant under the extended transformations $\deltae$. \footnote{See Appendix \ref{Exp} for a detailed explanation.}
Using (\ref{vare}), there follows that
\be
\deltas\Delta S_k^{(i)}=\int d^dx\, \epsilon\,k\frac{\delta}{\delta k}\Delta S_k^{(i)}
\qquad
\mathrm{for}\ i\in{T,gh}
\label{varcut}
\ee
where the functional variation with respect to $k$ acts only on the cutoffs $\cR_k$.

The case $i=\omega$ works a little differently, because $\omega$ does not 
transform homogeneously:
\be
\deltas\Delta S_k^\omega=\int d^dx\, \epsilon\,k\frac{\delta}{\delta k}\Delta S_k^\omega
-\frac{1}{2}\int d^dx\sqrt{\bg}\,\left(\epsilon\cR_k\omega+\omega r_{0}\epsilon\right)\ .
\label{varcutomega}
\ee
Thus this term is {\it not} invariant under $\deltae$.
\footnote{The analogous term in \cite{Percacci:2011uf} was invariant
because it was written in terms of the field $e^\omega$ that transforms homogeneously.}

%%%%%%%%%%%%%%%%%%%%%%%%%%%%%%%%%%%
\subsection{The modified Ward identity}\label{Wardid}
%%%%%%%%%%%%%%%%%%%%%%%%%%%%%%%%%%%

We now have all the ingredients that are needed to derive the
Ward identity for the SW tranformations $\deltas$.
One could follow step by step the derivation given in \cite{Percacci:2016arh},
which was based on the integro-differential equation satisfied by the EAA.
Alternatively, we follow here the logic of \cite{Morris:2016spn}.
We subject $W_k$ to a SW transformation,
with fixed sources and fixed $k$.
Since the actions $S$, $S_{GF}$ and $S_{gh}$ are invariant by construction,
the only variations come from the cutoff and source terms:
\be
\deltas W_k=
-\langle\deltas \Delta S_k^T\rangle
-\langle\deltas \Delta S_k^\omega\rangle
-\langle\deltas \Delta S_k^{gh}\rangle
-\int d^dx j\epsilon.
\ee
The variations of the cutoff terms have been given in (\ref{varcut},\ref{varcutomega}).
Their expectation values involve two- and one-point functions,
that we can reexpress in terms of connected two-point functions and one-point functions
as follows
\bea
&&
-\frac{1}{2}
\Tr
\int \epsilon k\frac{\delta \cR_k}{\delta k}\,
\frac{\delta^2 W_k}{\delta j_T\delta j_T}
-\frac{1}{2}\int d^dx\sqrt{\bg}\,
\frac{\delta W_k}{\delta j_T}
\int \epsilon k\frac{\delta \cR_k}{\delta k}\,
\frac{\delta W_k}{\delta j_T}
\nonumber
\\
&&
-\frac{1}{2}\Tr\,\int \epsilon k\frac{\delta \cR_k}{\delta k}\,
\frac{\delta^2 W_k}{\delta j\delta j}
+\frac{\delta W_k}{\delta j}
\int \epsilon k\frac{\delta \cR_k}{\delta k}\,
\frac{\delta W_k}{\delta j}
-\int d^dx \sqrt{\bg}\epsilon\cR_k\frac{\delta W_k}{\delta j}
\nonumber\\
&&
-\Tr\,\int \epsilon k\frac{\delta \cR_k}{\delta k}\,
\frac{\delta^2 W_k}{\delta J\delta J_*}
+2\frac{\delta W_k}{\delta J_*}
\int \epsilon k\frac{\delta \cR_k}{\delta k}\,
\frac{\delta W_k}{\delta J}
\nonumber
\eea
where we use the shorthand (\ref{abrid})
and we suppress indices for notational clarity.
The variation of the EAA can be computed inserting these variations in (\ref{eaa}).
The terms containing the sources cancel out,
as does the term linear in $\omega$ from (\ref{varcutomega})
and the variations of the cutoff terms evaluated on the classical fields.
There remain only the terms with the connected two-point functions,
that can be re-expressed in terms of the EAA:
\bea
\deltas \Gamma_k&=&
\frac{1}{2}
\Str\left(
\frac{\delta^2\Gamma_k}{\delta\phi\delta\phi}+\cR_k\right)^{-1}
\int \epsilon k\frac{\delta \cR_k}{\delta k}
\nonumber\\
&=&\frac{1}{2}
\Tr\left(
\frac{\delta^2\Gamma_k}{\delta h^T\delta h^T}+\cR_k\right)^{-1}
\int \epsilon k\frac{\delta \cR_k^T}{\delta k}
+\frac{1}{2}
\Tr\left(
\frac{\delta^2\Gamma_k}{\delta\omega\delta\omega}+\cR_k\right)^{-1}
\int \epsilon k\frac{\delta \cR_k}{\delta k}
\nonumber\\
&&
-\Tr\left(
\frac{\delta^2\Gamma_k}{\delta C^*\delta C}+\cR_k\right)^{-1}
\int \epsilon k\frac{\delta \cR_k}{\delta k}
+\ldots\ .
\label{wir}
\eea
Here we use the same superfield notation as in (\ref{lerge}),
and the ellipses indicate further mixing terms that arise in the inversion
of the Hessian.

Comparing (\ref{wir}) and (\ref{lerge}) we see that
\be
\deltas\Gamma_k=\int \epsilon k\frac{\delta \Gamma_k}{\delta k}\ ,
\label{main}
\ee
where we recall that the variation on the l.h.s.
involves only the field arguments of $\Gamma_k$
and leaves $k$ fixed.
We have thus arrived at a remarkably simple result:
with our choices for the gauge and cutoff terms,
the anomalous variation in the mSWWI
is given by the ``beta functional'' of the theory, as expressed
by the r.h.s. of the local ERGE.

%%%%%%%%%%%%%%%%%%%%%%%%%%%%%%%%%%%
\subsection{The reduced flow equation}
%%%%%%%%%%%%%%%%%%%%%%%%%%%%%%%%%%%

Recalling the definition (\ref{vare}), 
we can rewrite (\ref{main}) simply as
\be
\deltae\Gamma_k=0\ .
\ee

This is a statement of invariance of the EAA under a particular
realization of the Weyl group.
At the level of finite transformations
\be
\Gamma_k(h^{T\mu}{}_\nu,C^*_\mu,C^\mu,\omega;\bsigma,\hg_{\mu\nu})
=\Gamma_{\Omega^{-1}k}(h^{T\mu}{}_\nu,C^*_\mu,C^\mu,\omega-\log\Omega;
\bsigma+\log\Omega,\hg_{\mu\nu})\ .
\label{hat}
\ee
We can therefore rewrite the action entirely in terms of SW-invariant
variables.
Having chosen some of the fields to be invariant
obviously simplifies the task.
The choice of variables that we find both conceptually most satisfying 
and technically most useful is the following:
\be
\hk=e^\bsigma k\ ;\quad
h^{T\mu}{}_\nu\ ;\quad
C^*_\mu\ ;\quad
C^\mu\ ;\quad
\sigma=\bsigma+\omega\ ;\quad
\hg_{\mu\nu}\ .
\ee
In the spirit of Weyl's theory, we are using the background dilaton field $\bchi=e^{-\bsigma}$
as unit of length and measure everything in its units.
\footnote{We avoid the alternative definition $\hat k=e^{\omega}k$
used in \cite{Percacci:2016arh} because we find it awkward to have a dynamical variable in the cutoff scale.
Another possible invariant metric would be
$\tilde g_{\mu\nu}=e^{2\omega}\bg_{\mu\nu}$.
Note the relation between invariants:
$\tilde g_{\mu\nu}=e^{2\sigma}\hg_{\mu\nu}$.
The alternative definition $\hat{h}^T_{\mu\nu}=e^{2\omega}h^T_{\mu\nu}$
would lead to a more complicated (off-diagonal) Jacobian.}
The solution of the mSWWI is therefore a functional 
\be
\hat\Gamma_\hk(h^{T\mu}{}_\nu,C^*_\mu,C^\mu,\sigma;\hg_{\mu\nu})
=\Gamma_k(h^{T\mu}{}_\nu,C^*_\mu,C^\mu,\omega;\bsigma,\hg_{\mu\nu})\ .
\label{hateaa}
\ee
As expected the mSWWI eliminates the dependence of
the EAA on the dynamical variable $\omega$ and on the background variable $\bsigma$,
replacing them by the single invariant $\sigma$.
In the process one also has to redefine the background
as well as the cutoff.

We must now rewrite the flow equation in terms of the new variables.
Taking the total variation of both sides of (\ref{hat}),
regarded as functionals of all their arguments,
and comparing the coefficients of each differential,
one obtains the following transformation rules:
\bea
k\frac{\delta\Gamma}{\delta k}&=&\hk\frac{\delta\hat\Gamma}{\delta \hk}\ ;\qquad
\frac{\delta\Gamma}{\delta\hg_{\mu\nu}}=\frac{\delta\hat\Gamma}{\delta\hg_{\mu\nu}}\ ;\qquad
\frac{\delta\Gamma}{\delta\bsigma}=
\frac{\delta\hat\Gamma}{\delta\sigma}
+\hk \frac{\delta\hat\Gamma}{\delta \hk}
\nonumber\\
\frac{\delta\Gamma}{\delta h^{T\mu}{}_\nu}&=&
\frac{\delta\hat\Gamma}{\delta h^{T\mu}{}_\nu}\ ;\qquad
\frac{\delta\Gamma}{\delta\omega}=\frac{\delta\hat\Gamma}{\delta\sigma}\ ;\qquad
\frac{\delta\Gamma}{\delta C^*_\mu}=\frac{\delta\hat\Gamma}{\delta C^*_\mu}\ ;\qquad
\frac{\delta\Gamma}{\delta C^\mu}=\frac{\delta\hat\Gamma}{\delta C^\mu}\ .
\eea
The reduced flow equation for the functional $\hat\Gamma$ has the form
$\hk\frac{\delta\hat\Gamma}{\delta\hk}=\ldots$.
Its r.h.s. is the r.h.s. of (\ref{lerge}), that we must rewrite in terms of the new variables.
In terms of the invariant ``superfield'' 
$\hat\phi=(h^{T\mu}{}_\nu,\sigma,C^*_\mu,C_\mu)$,
one obtains
\bea
\hk\frac{\delta\hat\Gamma_\hk}{\delta \hk}&=&
\frac{1}{2}
\Str\left(
\frac{\delta^2\hat\Gamma_\hk}{\delta\hat\phi\delta\hat\phi}+\hat\cR_\hk\right)^{-1}
\hk\frac{\delta\hat\cR_\hk}{\delta\hk}
\nonumber\\
&=&\frac{1}{2}
\Tr\left(
\frac{\delta^2\hat\Gamma_\hk}{\delta\hat h^T\delta\hat h^T}+\hat\cR_\hk\right)^{-1}
\hk\frac{\delta \hat\cR_\hk^T}{\delta\hk}
+\frac{1}{2}
\Tr\left(
\frac{\delta^2\hat\Gamma_\hk}{\delta\sigma\delta\sigma}+\hat\cR_\hk\right)^{-1}
\hk\frac{\delta \hat\cR_\hk}{\delta\hk}
\nonumber\\
&&
-\Tr\left(
\frac{\delta^2\hat\Gamma_\hk}{\delta C^*\delta C}+\hat\cR_\hk\right)^{-1}
\hk\frac{\delta \hat\cR_\hk}{\delta\hk}
+\ldots\ .
\label{rerge}
\eea
We see that the reduced flow equation has exactly the same form of the original one,
except for being formulated in terms of invariant variables.

We note that this equation could be derived by rewriting the cutoff action as a functional the new variables:
\bea
\Delta\hat S_\hk^T(h^{T\mu}{}_\nu;\hg_{\mu\nu})
&=&\Delta S_k(h^{T\mu}{}_\nu;\hg_{\mu\nu},\bsigma)
=\frac{1}{2}\int d^dx\sqrt{\hg}\,
h^{T\mu}{}_\nu\,\hat\cR_\hk(\hDelta) h^{T\nu}{}_\mu
\nonumber\\
\Delta\hat S_\hk^\sigma(\sigma;\hg_{\mu\nu})
&=&\Delta S_k^\omega(\omega;\hg_{\mu\nu},\bsigma)
=\frac{1}{2}\int d^dx\sqrt{\hg}\,\omega\,\hat\cR_\hk(\hDelta)\omega\ ,
\nonumber\\
\Delta\hat S_\hk^{gh}(C^*_\mu,C^\mu;\hg_{\mu\nu})
&=&\Delta S_k(C^*_\mu,C^\mu;\hg_{\mu\nu},\bsigma)
=\int d^dx\sqrt{\hg}\,
C_\mu^* \hg^{\mu\nu}\hat\cR_\hk(\hDelta) \hat C_\nu\ ,
\eea
where now 
\be
\hat\cR_\hk^{(i)}(\hDelta)=\,u^d r(y)%=c_i \,\bchi^d u^d r(y)
\ , 
\qquad y=\frac{1}{\hk^2}\hDelta\ ,
\qquad
\hDelta=\frac{1}{\bchi^2}\bDelta=\hg^{\mu\nu}\cD_\mu\cD_\nu
\ ,\qquad i\in\{T,\omega,gh\}
\ee

%%%%%%%%%%%%%%%%%%%%%%%%%%%%%%%%%%%%%%%%%%%%%%%%%%
\subsection{The global form of the reduced ERGE}

In the discussion, we have introduced the
extended transformations where besides the fields,
also the cutoff is subjected to Weyl transformation.
This is natural from the point of view of dimensional analysis,
but it leads to the consequence that the cutoff cannot be
regarded as constant anymore.
The ERGE can be easily generalized to the case of non-constant cutoff,
but its physical interpretation becomes then less clear.
The flow of the FRGE in theory space would depend on a function, instead of a single parameter,
which would be somewhat reminiscent of the ``many-fingered time''
of General Relativity.
It would be interesting to explore a possible connection of the local ERGE with the notion of local RG \cite{Osborn:1991mk,Osborn:1991gm},
which would then give it a direct physical meaning.
We refrain from doing so here.
Instead, we have noted that the solution of the mSWWI
implies that also the cutoff has to be replaced,
as an argument of the EAA, by the quantity $\hk$.
Unlike $k$, it is invariant under (extended) SW transformations.
It is therefore consistent to assume that 
\be
\hk=k e^\bsigma=\mathrm{constant}
\label{condition}
\ee
If $\hk$ is constant, we can replace
$$
\hk(x)\frac{\delta\hat\Gamma_\hk}{\delta \hk(x)}\qquad \mathrm{by}\qquad
\hk\frac{d\hat\Gamma_\hk}{d \hk}
$$
and the reduced ERGE becomes again an ordinary differential equation,
whose solution are curves in theory space depending on the
single parameter $\hk$.
In this way the local ERGE can be seen just as an intermediate mathematical construction.

%%%%%%%%%%%%%%%%%%%%
\section{Discussion}
%%%%%%%%%%%%%%%%%%%%

Let us summarize the main steps of our procedure.
We started from the exponential parametrization (\ref{exppar})
and demanded that physical results should not change under
the SW transformations (\ref{ennio}), (\ref{varbsigma}).
This is part of the requirement of background-independence.
The classical action is trivially invariant under these transformations,
because it is formulated in terms of a single metric,
but the quantum effective action cannot be.
In particular, the EAA, which is an effective action depending on an external
cutoff scale $k$, cannot be invariant, because the gauge-fixing term
and even more importantly the cutoff term are not.
There is therefore a kind of anomaly.
By making certain choices, we have however been able to define the EAA
in such a way that the {\it only} source of non-invariance is the presence of the
cutoff $k$.
One can then extend the definition of the SW transformations
by also transforming $k$, which henceforth had been kept fixed.
The natural transformation is (\ref{deltae}),
on account of the dimensionality of $k$.
This implies that $k$ cannot be assumed to be constant.
We have therefore generalized the ERGE by allowing the cutoff to be a positive function
on spacetime.
The resulting local ERGE has the same form as the usual one,
except for the appearance of functional derivatives with respect to $k$
in place of ordinary derivatives.
One finds that the r.h.s. of the mSWWI
is identical to the r.h.s. of the local ERGE.
Then, the mSWWI just expresses the fact that the EAA is {\it invariant}
under the extended transformations.
This is the central result of the present paper.
It establishes that the mSWWI and the flow equation are manifestly compatible
and can be solved simultaneously.
The solution of the mSWWI consists in writing the EAA as a functional $\hat\Gamma$
of a new set of variables that are invariant under the extended SW transformations,
as in (\ref{hateaa}).
We have then shown that the functional $\hat\Gamma$ satisfies a local ERGE (\ref{rerge})
that is formally identical to the one satisfied by the original EAA.

The main aim of this work was to reduce the number of independent
arguments that the EAA depends on, when the background field
method is used.
Let us discuss how this goal has been achieved.
In the case of CORE gravity the fiducial metric is always held fixed,
so the EAA can be seen as a functional $\Gamma_k(\omega;\bsigma)$,
where $\bsigma$ is the background field (the conformal factor of
the background metric) and $\omega$ is the quantum field
(such that $\bsigma+\omega$ is the conformal factor
of the full dynamical metric).
The mSWWI (\ref{COREmSWWI}) shows that the EAA can be
rewritten in terms of a functional of $\sigma$ alone,
thereby reducing the number of scalar fields that it depends on
from two to one, as desired. 

Let us now see how the counting works in full gravity.
In principle, we begin from an EAA
$\Gamma_k(h^{T\mu}{}_\nu,\omega;\bg_{\mu\nu})$,
depending on 9+1+10=20 functions, instead of the desired 10
(we do not count here the ghosts, which are irrelevant for this discussion).
We wanted to reduce this number by one by solving the mSWWI.
However, in order to apply our techniques, we had to
reparametrize the background metric by splitting off
its conformal part as in (\ref{bgexp}).
This can only be done by first choosing a reference metric
$\hg_{\mu\nu}$ in the same conformal class as the background.
In this way the EAA has become a functional
$\Gamma_k(h^{T\mu}{}_\nu,\omega;\bsigma,\hg_{\mu\nu})$
depending on 9+1+1+10=21 functions.
The solution of the mSWWI (\ref{main}) allows us to
rewrite the EAA as a functional
$\hat\Gamma_\hk(h^{T\mu}{}_\nu,\sigma;\hg_{\mu\nu})$
where, just as in the CORE case, the two functions $\omega$ and $\bsigma$ have been replaced by the single function $\sigma$.
But now this depends again on 9+1+10 functions:
it appears that we have merely traded the 
original dependence on the background metric
by the dependence on the fiducial metric.

Why is full gravity different from CORE gravity?
The only difference is that in the discussion of
full gravity we keep track of the dependence on the fiducial metric, whereas in CORE gravity this is ignored.
In fact the same issue would be present also in CORE gravity
if we took into account the dependence on the fiducial metric.

In both cases, the problem is that the fiducial metric $\hg_{\mu\nu}$
is another artificial choice that enters in the definition of the EAA,
just like the background metric, 
and no physical result should depend on it.
The two fields $\bsigma,\hg_{\mu\nu}$ would count as 10 independent
variable if $\Gamma_k$ was invariant under FW transformations.
This invariance, however, is broken by the cutoffs.
Alternatively, we could at least try to solve the corresponding modified FW WI.
Unfortunately, as seen in Eq. (\ref{deltafw}), the transformation
of the cutoff for $\omega$ under such FW transformations is very complicated and there is
no hope of achieving a practical solution. 
In addition, in full gravity the transformation of the other cutoffs 
is similarly intractable
\footnote{We observe that if we used the cutoff (\ref{firstcut}),
as in \cite{dm3}, invariance under FW transformations would be trivial.
In that case we would not have been able to solve the mSWWI, though.}
\be
\deltaf\Delta S^{T}_{k}=-\frac{1}{2}\int d^{d}x\sqrt{\bg}h^{T\mu}{}_{\nu}\left[\partial_{\lambda}\epsilon\frac{\partial \cR^{T}_{k}}{\partial(\partial_{\lambda}\bsigma)}+\partial_\lambda\partial_\alpha\epsilon\frac{\partial \cR^{T}_{k}}{\partial(\partial_\lambda\partial_\alpha\bsigma)}+\cdots\right]h^{T\nu}{}_{\mu},
\ee
\be
\deltaf\Delta S^{gh}_{k}=-\int d^{d}x\sqrt{\bg}C^{*}_{\mu}\left[\partial_{\lambda}\epsilon\frac{\partial \cR^{gh}_{k}}{\partial(\partial_{\lambda}\bsigma)}+\partial_\lambda\partial_\alpha\epsilon\frac{\partial \cR^{gh}_{k}}{\partial(\partial_\lambda\partial_\alpha\bsigma)}+\cdots\right]C^{\mu}.
\ee
Thus, with the chosen cutoff, there is no way to solve exactly
the mFWWI.
Additional approximations may allow one to do so.
We show in Appendix (\ref{mFWWI}) how to obtain a simple
and exactly solvable mFWWI by using a different type of cutoff.
In that case it is the mSWWI that is too complicated to solve.

Thus we conclude that the task of reducing the
number of functions in the EAA by one cannot be solved
by the methods used here in generality.
If there was any separate physical argument selecting a preferred
metric $\hg_{\mu\nu}$ within its conformal class,
then the methods discussed here would provide the desired
reduction of the independent variables.
This solution could probably be extended to the full
$\bg_{\mu\nu}$-dependence of the EAA (as opposed to just its
conformal factor) by using the $GL(4)$-invariant formulation
discussed in \cite{mean,Percacci:2009ij}.
The methods proposed are not restricted to the
Wetterich equation but can be extended also
to proper time-type flow equations,
both approximate \cite{flop2} and exact \cite{dealwis}.

\bigskip

\noindent
{\bf Acknowledgements} R.P. wishes to thank T. Morris
for a useful correspondence.

%%%%%%%%%%%%%%%%%%%%%%%%
\appendix
%%%%%%%%%%%%%%%%%%%%%%%%

\section{Weyl calculus}\label{weylcal}

The way to preserve background Weyl invariance in the quantum theory 
has been discussed in \cite{englert,fv,flop,shapo,Pagani:2013fca}.
In \cite{Percacci:2011uf} this relied on the existence of
a scalar field $\chi$ called the dilaton.
Here we do not need to appeal to the existence of an additional
degree of freedom, but use instead the inverse square root of the 
conformal factor of the background metric $\bchi=e^{-\bsigma}$. 
It transforms under Weyl transformations as
\be
\delta\bchi=-\epsilon\bchi\ ,
\ee
hence it can be identified with the background value of a dilaton.
We can use $\bchi$ to construct a pure-gauge abelian gauge field
$\kappa_\mu=-\bchi^{-1}\partial_\mu\bchi=\partial_\mu\bsigma$, transforming under 
Weyl transformations as
\be
\delta\kappa_\mu=\partial_\mu\epsilon\ .
\ee
Let $\bnabla_\mu$ be the covariant derivative with respect to the
Levi-Civita connection of the metric $\bg$ and
$\hnabla_\mu$ be the covariant derivative with respect to the
Levi-Civita connection of the metric $\hg$.
They are related by
\be
\hat\Gamma_\mu{}^\lambda{}_\nu=
\bar\Gamma_\mu{}^\lambda{}_\nu-
\delta^\lambda_\mu\kappa_\nu-\delta^\lambda_\nu\kappa_\mu
+\bg_{\mu\nu}\bg^{\lambda\tau}\kappa_\tau\ .
\ee
The connection coefficients $\hat\Gamma$ are invariant under 
background Weyl transformations, as is obvious since
the metric $\hg$ is.
We say that a tensor $t$ has weight $\alpha$ if it transforms
under background Weyl transformation as
\be
\delta t=\alpha\,\epsilon\, t\ .
\ee
(Here we do not write tensor indices, as they are the same
on both sides of the equation.)
For example, the background metric has weight 2,
as does the the fluctuation $h_{\mu\nu}^T$.
For any tensor $t$ of weight $\alpha$, we
define the Weyl-covariant derivative as
\be
\label{wcovder}
\cD_\mu t=\hnabla_\mu t-\alpha\kappa_\mu t\ .
\ee
It is a tensor with the same weight as $t$.
We note in particular the special cases
\be
\cD_\rho \bg_{\mu\nu}=0\ ;\qquad\cD_\rho\bchi=0.
\ee
The fields $\bsigma$ and $\omega$ transform inhomogeneously
and therefore have to be treated separately.
Their Weyl-covariant derivatives are defined as
\be
\cD_\rho\bsigma=\partial_\rho\bsigma-\kappa_\rho=0\ ;\qquad
\cD_\rho\omega=\partial_\rho\omega+\kappa_\rho
\ee
and are {\it in}variant (reflecting the absence of a homogeneous
term in their transformation).

%%%%%%%%%%%%%%%%%%%%%%%%%%%%%%%%%%%
\section{The Local Exact Renormalization Group Equation}\label{localap}
%%%%%%%%%%%%%%%%%%%%%%%%%%%%%%%%%%%

In this Appendix, we derive a renormalization group equation for theories containing an $x$-dependent scale $k$. We obtain such an equation for a general field $\phi$ such that the result can be applied to any theory. 
We start with the generating functional of connected Green functions
\be
e^{W_k(j)}
=
\int (\mathcal{D}\phi)
Exp\Big[-S(\phi)
-\Delta S_k(\phi)+\int d^dx \left(j\phi\right)\Big]
\ee

The EAA (\ref{eaa}) is therefore a {\it functional} of $k$.
We can calculate the variation of $\Gamma_k$ under an infinitesimal
change in the cutoff function.
As usual one starts from varying $W_k$, to obtain
\be
\int\delta k\frac{\delta W_k}{\delta k}
=-\left\langle\int\delta k\frac{\delta \Delta S_k}{\delta k}\right\rangle
=-\frac{1}{2}\Tr\langle \phi\phi\rangle\int \delta k\frac{\delta \cR_k}{\delta k},
\label{varw}
\ee
where we use the notation 
\be
\int f \delta k\frac{\delta }{\delta k}=\int d^dx\, f(x)\delta k(x)\frac{\delta}{\delta k(x)}\ .
\label{abrid}
\ee
The calculation then follows closely the derivation of the Wetterich equation,
except for the fact that $\delta k$ remains inside the traces.
One obtains
\be
\int \delta k\frac{\delta\Gamma_k}{\delta k}=\frac{1}{2}
\Tr\left(
\frac{\delta^2\Gamma_k}{\delta\phi\delta\phi}+\cR_k\right)^{-1}
\int \delta k\frac{\delta \cR_k}{\delta k}.
\ee
Since $\delta k$ is arbitrary, we obtain a local flow equation
giving $\frac{\delta\Gamma_k}{\delta k(x)}$ by simply removing
the integrals and the factors $\delta k$ from both sides.
In the case when $k$ is constant the functional derivatives 
reduce to ordinary derivatives and the local ERGE reduces to the standard ERGE.

In the case of gravity, the flow equation would read
\bea
\int \delta k\frac{\delta\Gamma_k}{\delta k}
&=&
\frac{1}{2}
\Str\left(
\frac{\delta^2\Gamma_k}{\delta\phi\delta\phi}+\cR_k\right)^{-1}
\int \delta k\frac{\delta \cR_k}{\delta k}
\nonumber\\
&=&
\frac{1}{2}
\Tr\left(
\frac{\delta^2\Gamma_k}{\delta h^T\delta h^T}+\cR_k\right)^{-1}
\int \delta k\frac{\delta \cR_k}{\delta k}
+\frac{1}{2}
\Tr\left(
\frac{\delta^2\Gamma_k}{\delta\omega\delta\omega}+\cR_k\right)^{-1}
\int \delta k\frac{\delta \cR_k}{\delta k}
\nonumber\\
&&
-\Tr\left(
\frac{\delta^2\Gamma_k}{\delta C^*\delta C}+\cR_k\right)^{-1}
\int \delta k\frac{\delta \cR_k}{\delta k}
+\ldots\ .
\label{lerge}
\eea
In the first line we have written the equation in terms of the ``superfield''
$\phi=(h^T_{\mu\nu},\omega,C^*_\mu,C_\mu)$
and $\cR_k$ is a block-diagonal matrix.
In the second line the supertrace has been expanded, neglecting
off-diagonal terms, which are denoted by the ellipses.

%%%%%%%%%%%%%%%%%%%%%%%%%%%%%%%%%%%%%%%%%%%%
\section{Other CORE cutoffs}\label{Cutoffs}
%%%%%%%%%%%%%%%%%%%%%%%%%%%%%%%%%%%%%%%%%%%

In this Appendix, we discuss other possible cutoffs in the framework of CORE gravity. This is done for completeness since the main purpose of the paper was to study background invariance. We start by considering a cutoff constructed with the Laplacian of the metric $\hg$, as it is done in \cite{machado}. That is, we use $\hat\Delta=\hg^{\mu\nu}\hat\nabla_{\mu}\hat\nabla_{\nu}$ and define 
\be
    \hat\Delta S_k=\frac{1}{2}\int d^dx \sqrt{\hg}\omega \cR_k(-\hat\Delta) \omega
    \label{Three}
\ee
The transformations $\deltaf$, $\deltas$ and $\deltab$ discussed in Section (2.1) produce
\bea
\deltaf\hat\Delta S_k&=&\frac{1}{2}\int d^{d}x \sqrt{\hg}(d\epsilon)\omega \cR_{k}\omega-\frac{1}{2}\int d^{d}x\sqrt{\hg}\omega\left[\epsilon\frac{\partial \cR_{k}}{\partial\bsigma}+\partial_{\mu}\epsilon\frac{\partial R_{k}}{\partial(\partial_{\mu}\bsigma)}+\ldots\right]\omega,
\\
\deltas\hat\Delta S_k&=&-\frac{1}{2}\int d^{d}x\sqrt{\hg}(\epsilon \cR_{k}\omega+\omega \cR_{k}\epsilon),
\\
\deltab\hat\Delta S_k&=&0.
\eea
We see that it is not possible to find simple transformation property for all the set $\deltaf$, $\deltas$, $\deltab$. In particular, the transformation under $\deltaf$ contains an infinite number of terms which generates a complicated transformation for $\Gamma_{k}$. 

If, instead, we use a covariant derivative, we find a useful transformation under $\delta^{(F)}$ but spoil the other two as we show now. 
We define a slightly different covariant derivative whose structure is similar to the one given in Appendix \ref{weylcal}. If the variation of a tensor $t$ under $\delta^{(F)}$ is $\delta^{(F)}=\beta\epsilon t$, we define the covariant derivative $\cD'_{\mu}$ such that $\cD'_{\mu}t=\bnabla t+\beta\partial_{\mu}\bsigma t$
(note the different sign relative to $\cD$, which is due to
the different transformation of $\bsigma$). 

In particular, for  $\omega$ we have $\cD'_{\mu}\omega=\partial_{\mu}\omega$. Then, introducing the operator $\hat\Delta^{F}=\hg^{\mu\nu}\cD'_{\mu}\cD'_{\nu}$ the new cutoff is
\be
    \hat\Delta S^{(1)}_k=\frac{1}{2}\int d^dx \sqrt{\hg}\omega \cR_k(-\hat\Delta^{F}) \omega
    \label{Four}
\ee
whose transformations are
\bea
\deltaf\hat\Delta S^{(1)}_k&=&\int d^dx\epsilon k\frac{\delta}{\delta k}\hat\Delta S^{(1)}_k,
\\
\delta^{(S)}\hat\Delta S^{(1)}_k&=&-\frac{1}{2}\int d^{d}x\sqrt{\hg}(\epsilon \cR_{k}\omega+\omega \cR_{k}\epsilon)
\nonumber\\
&&+\frac{1}{2}\int d^{d}x\sqrt{\hg}\omega\left[\partial_{\mu}\epsilon\frac{\partial \cR_{k}}{\partial(\partial_{\mu}\bsigma)}+\partial_\mu\partial_\nu\epsilon\frac{\partial \cR_{k}}{\partial(\partial_\mu\partial_\nu\bsigma)}+\ldots\right]\omega,
\\
\delta^{(B)}\hat\Delta S^{(1)}_k&=&\frac{1}{2}\int d^{d}x\sqrt{\hg}\omega\left[\partial_{\mu}\epsilon\frac{\partial \cR_{k}}{\partial(\partial_{\mu}\bsigma)}+\partial_\mu\partial_\nu\epsilon\frac{\partial \cR_{k}}{\partial(\partial_\mu\partial_\nu\bsigma)}+\ldots\right]\omega.
\eea
As expected, the transformation under $\delta^{(F)}$ has a compact form.

Finally, we introduce the operator $\bar\Delta^{B}=-\bg^{\mu\nu}\cD_{\mu}\cD_{\nu}$,
which differs from the Laplacian introduced in section 2.2
because the weight of the fields may differ.
In particular, here $\cD_{\mu}\omega=\partial_{\mu}\omega$. 
The cutoff is then given by
\be
    \bar\Delta S^{(B)}_k=\frac{1}{2}\int d^dx \sqrt{\bg}\omega \cR_k(-\bar\Delta^{B}) \omega,
    \label{Five}
\ee
and the transformations by
\bea
    \delta^{(F)}\Delta S^{(B)}_{k}&=&-\frac{1}{2}\int d^{d}x\sqrt{\bg}\omega\left[\partial_{\mu}\epsilon\frac{\partial \cR_{k}}{\partial(\partial_{\mu}\bsigma)}+\partial_\mu\partial_\nu\epsilon\frac{\partial \cR_{k}}{\partial(\partial_\mu\partial_\nu\bsigma)}+\ldots\right]\omega,
\\
    \delta^{(S)}\Delta S^{(B)}_{k}&=&\int d^dx\epsilon k\frac{\delta}{\delta k}\Delta S^{(B)}_{k}-\frac{1}{2}\int d^{d}x\sqrt{\bg}(\epsilon \cR_{k}\omega+\omega \cR_{k}\epsilon),
\\
    \delta^{(B)}\Delta S^{(B)}_{k}&=&\int d^dx\epsilon k\frac{\delta}{\delta k}\Delta S^{(B)}_{k}.
\eea

We can derive a Ward identity for each transformation and cutoff. The inclusion of the last two cutoffs with the covariant derivatives would lead to Ward identities similar to (\ref{COREmSWWI}), with the change of $\deltas$ by $\deltaf$ and $\deltab$ respectively. We stress again that a CORE theory with $\deltab$-invariance would be trivial so it is not worth studying it. Thus, we end up with $\deltas$ and $\deltaf$, and we realize that there is no cutoff/effective action satisfying simple Ward identities for both transformations. According to our interest, we choose one or another. In the main part of the paper we took $\cR_{k}(\bDelta^{W})$ because in this case a satisfactory extension to full gravity was found, as seen in Section \ref{Full}.

\section{The modified Fiducial Weyl Ward Identity}\label{mFWWI}

In this Appendix, we discuss briefly how a simple WI for the FW transformation in full gravity can be obtained, by using 
a suitable cutoff. The steps are similar to the ones followed in Section \ref{Full}. We start by writing the gauge-fixing term which will have a similar form than before: 
\be
S_{GF}=\frac{1}{2\alpha}\int d^d x \sqrt{\hg}\,F_\mu Y^{\mu\nu} F_\nu,
\ee
where now we use the covariant derivative $\cD'$ defined in the Appendix \ref{Cutoffs}
\be
F_\mu=\cD'_\rho h^\rho{}_\mu-2(\b+1)\cD'_\mu \omega\ .
\ee
On the other hand, it is useful to define a new $Y$
\be
Y^{\mu\nu}=e^{-(d-2)\bsigma}\hg^{\mu\nu}\ .
\ee
The deduction of the ghost action will follow the standard steps, and its final form will be
\be
S_{gh}(C^*_\mu,C_\mu;\bg_{\mu\nu},\bsigma)=\int d^dx\sqrt{\hg}\,
C^*_\mu Y^{\mu\nu}\Delta_{FP\nu\rho} C^\rho \ 
\ee
where 
\be
\Delta_{FP\mu\nu} C^\nu=
\cD'_\rho\left(
(\delta^{(Q)}_C \mX)^\rho{}_\mu
-\frac{1+\beta}{d}\delta^\rho{}_\mu
\tr(\delta^{(Q)}_C \mX)
\right).
\ee
Finally, we introduce the cutoffs for all the fluctuations
\bea
\Delta S_k^T(h^T;\hg,\bsigma)&=&
\frac{1}{2}\int d^dx\sqrt{\hg}\,
h^{T\mu}{}_\nu\cR_k(\hat \Delta^{F}) h^{T\nu}{}_\mu\ ,
\nonumber\\
\Delta S_k^\omega(\omega;\hg,\bsigma)&=&
\frac{1}{2}\int d^dx\sqrt{\hg}\,\omega\,\cR_k(\hat \Delta^{F})\omega\ ,
\nonumber\\
\Delta S^{gh}_k(C^*,C;\hg,\bsigma)&=&
\int d^dx\sqrt{\hg}\,
C_\mu^* \cR_k(\hat \Delta^{F}) C^\mu\ .
\eea
Note that the only difference with respect to Eq. (\ref{allgencutoff}) is in the volume element and the argument of the cutoff functions $\cR^{i}_{k}$. Promoting $k$ to be a function of the spacetime and following the lines of Subsection \ref{Wardid}, we obtain
\be
\deltaf\Delta S_k^{(i)}=\int d^dx\, \epsilon\,k\frac{\delta}{\delta k}\Delta S_k^{(i)}
\qquad
\mathrm{for}\ i\in{T,gh,\omega}.
\ee
Now, taking into account that $S_{GF}$ and $S_{gh}$ are invariant under $\deltaf$, it is not difficult to realize that
\be
\deltaf\Gamma_k=\int \epsilon k\frac{\delta \Gamma_k}{\delta k}\ .
\ee
Therefore, the solution of the mFWWI is a functional
\be
\hat\Gamma_\hk(h^{T\mu}{}_\nu,C^*_\mu,C^\mu,\omega;\bg_{\mu\nu})
=\Gamma_k(h^{T\mu}{}_\nu,C^*_\mu,C^\mu,\omega;\bsigma,\hg_{\mu\nu})\ ,
\ee
where, as before, $\bg_{\mu\nu}=\hg_{\mu\nu}e^{2\bsigma}$ and $\hk=e^{\bsigma}k$. Thus, we see that the role of the mFWWI is to combine the functions $\hg_{\mu\nu}$, $\bsigma$ into only one function $\bg_{\mu\nu}$.

\section{Expansions}\label{Exp}

In this Appendix, we want to show how the function $r(y)$, acting on a tensor $t$ of weight $\alpha$, transforms under $\deltae$. The result can be straightforwardly extended to the field $\omega$. As usual, we start by expanding $r(y)$ in a Taylor series
\be
r(y)=\sum_{n=0}^\infty r_n y^n\ .
\ee
Then, we act on $t$ and find how each term of $r(y)t$ in the series transforms. The thing to bear in mind is that $y$ contains the covariant derivative $\cD_{\mu}$ which acts on all the fields including $k(x)$. The variation of the first term gives $\deltae(r_{0}t)=\alpha\epsilon r_{0}t$. Since $\deltae$ goes through the covariant derivative, the variation of the second term gives
\be
    \deltae(yt)=\alpha\epsilon(yt).
\ee
From this expression, we obtain how $y$ transforms. Since $\deltae(yt)=(\deltae y)t+y(\deltae t)$, we get that
\be
    \deltae y=\alpha[\epsilon,y].
\ee
When this last result is applied to the third term in the series, we get
\be
    \deltae (y^{2}t)=\deltae y(yt)+y(\deltae yt)=\alpha[\varepsilon,y]yt+y(\alpha\epsilon yt)=\alpha\epsilon y^{2}t.
\ee
If we proceed by induction, we arrive at
\be
    \deltae (y^{n}t)=\alpha\epsilon(y^{n}t).
\ee

Thus, we realize that $\deltae (r(y)t)=\alpha\epsilon r(y)t$. That is, $r(y)$ maps a tensor of weight $\alpha$ to another tensor of weight $\alpha$ under $\deltae$. Consequently, when analysing the variation of $\Delta S_{k}$ under $ \deltae$, we take the transformation of $r(y)t$ as just the transformation property of $t$. If we want to extend this result to the field $\omega$, we have to take into account that it transforms by a shift and that $\cD_{\mu}\omega$ is invariant under $\deltae$. Therefore, in this case we have
\be
    \deltae r(y)\omega=-r_{0}\epsilon.
\ee
These explicit outcomes lead us to the following conclusions
\be
    \deltae \Delta S^{T}_{k}=0, \ \  \deltae \Delta S^{gh}_{k}=0, \ \  \deltae\Delta S^{\omega}_{k}=-\frac{1}{2}\int d^dx\sqrt{\bg}\,\left(\epsilon\cR_k\omega+\omega r_{0}\epsilon\right). 
\ee
which are used in Section (3.3) to compute the transformation properties of the cutoffs for $h^{T}$, $C$ and $\omega$. Note that a crucial cancellation occurs between the transformation of the volume element and the transformation of $k^{d}$ in $\cR_{k}$.

%%%%%%%%%%%%%%%%%%%%%%%%%%%%%%%%%

\end{document}